\magnification=1200 \baselineskip=13pt \hsize=16.5 true cm \vsize=23 true cm
\def\parG{\vskip 10pt} \font\bbold=cmbx10 scaled\magstep2

{\bbold Broad Histogram Method}\parG
P.M.C. de Oliveira, T.J.P. Penna\par
Instituto de F\'\i sica, Universidade Federal Fluminense\par
av. Litor\^anea s/n, Boa Viagem, Niter\'oi RJ, Brazil 24210-340\parG
and H.J. Herrmann\par
ICA 1, University of Stuttgart, Pfaffenwaldring 27, 50569 Stuttgart, 
Germany\parG
e-mail PMCO@IF.UFF.BR
\vskip 0.3cm {\bbold Brazilian Journal of Physics 26, 677 (1996)}

\vskip 0.5cm \centerline{\bf Abstract}\parG
\leftskip=1cm \rightskip=1cm
Monte Carlo simulations of thermodynamic models are usually performed
according to Boltzmann's canonical distribution, with a fixed temperature
$T$. This can be very time consuming on the computer since one needs to do a
new computer run for each value of $T$. Salzburg {\sl et al} invented and
Ferrenberg and Swendsen perfected the histogram method to circumvent this, by
measuring the Boltzmann probability distribution as a function of the energy
$E$, for a fixed $T$. Instead of repeating the simulations for another value
$T'$, the measured probability distribution is simply reweighted through
analytic manipulations of the Boltzmann formula. The simulated Boltzmann
probability distribution has, however, exponentially decaying tails.
Therefore the statistics are poor away from the peak which is centered around
the average energy. Since the peak of the new $T'$ is centered somewhere on
these tails, the histogram method only works well for very small system
sizes.\par

We present a quite different approach, defining a non-biased random walk
along the $E$ axis with long range power-law decaying tails, and measuring
directly the degeneracy $g(E)$, without thermodynamic constraints. Our
arguments are general (model independent), and the method is shown to be
exact for the 1D Ising ferromagnet. Also for the 2D Ising ferromagnet, our
numerical results for different thermodynamic quantities agree quite well
with exact expressions.\par
\leftskip=0pt \rightskip=0pt

\vskip 0.5cm 
\centerline{\bf I) Introduction}\parG
The Monte Carlo approach is a fundamental tool to study the thermodynamic
properties of model systems [1]. Instead of taking into account all possible
states of the system, thermal averages are performed among a finite set of
states. These states form a random Markovian sequence generated according to
a dynamic rule which has as attractor fixed point the canonical Boltzmann
probability distribution

$$P_T(E) = {1\over Z_T} g(E) \exp(-E/T)
\eqno(1)$$

\noindent for each possible energy value $E$, where $T$ is the {\bf fixed}
temperature, $g(E)$ is the degeneracy of energy level $E$, and

$$Z_T = \sum_E g(E) \exp(-E/T)
\eqno(2)$$

\noindent is the partition function. Note that $E$ corresponds to the total
energy, and we have taken the Boltzmann constant $k_B = 1$. There are many
different dynamic rules obeying this probability distribution - the first one
being introduced in reference [2]. According to this rule one tries to make
some random movement in phase space, for instance through a one-spin flip,
starting from the current state of the system. If this movement leads to a
decrease of the energy, it is performed. If, however, an energy increment
$\Delta{E}$ would result from this movement, it is only performed  with
probability $\exp(-\Delta{E}/T)$. By repeating this rule many times, one
forms the quoted Markovian sequence of states, the thermal average $<Q>_T$ of
some quantity $Q$ (magnetization, susceptibility, specific heat, etc.) is
then simply the arithmetic mean of this quantity over the visited states. Of
course, one must take care of statistical correlations and fluctuations, in
order to get accurate values. There are many standard procedures [1] to do
this.

Normally one needs to calculate $<Q>_T$ as a function of $T$. So, one is
forced to repeat the entire procedure described in the last paragraph for
each different value of $T$. In order to save computer time, an appealing
strategy~[3] consists in extracting out the $T$ dependence from equations (1)
and (2). Note that $T$ only appears in the Boltzmann weight exponents, making
this task easy. First, the distribution $P_T(E)$ itself is measured by
accumulating in a histogram the number of visits to each value of $E$, during
the Markovian sequence of simulated states. Then, one can infer another
distribution $P_{T'}(E)$ corresponding to a different value $T'$ {\bf
without} performing any further computer run, simply by reweighting equations
(1) and (2). This approach is known as the histogram method, and has been
popularized by ref. [4]. In order to obtain the average $<Q>_T$, one needs to
accumulate also in another histogram the measured values of $Q$ corresponding
to each energy $E$. The thermal average at temperature $T$ is then

$$<Q>_T = \sum_E <Q(E)> P_T(E)\,\,\,\,\, ,
\eqno(3)$$

\noindent where $<Q(E)>$ means the average value of $Q$ obtained at fixed
energy $E$, i.e. the microcanonical average. Once one has the reweighted
distribution $P_{T'}(E)$, equation (3) can be applied to obtain $<Q>_{T'}$
for other (not simulated) temperatures $T'$.

The probability distribution $P_T(E)$ presents a sharp peak at $<E>_T$ and
decays exponentially on both sides. The larger the system size, the narrower
this peak. Thus, the computer measured $P_T(E)$ is only reliable around the
peak, the statistics being poor in the exponential tails. The reweighting
procedure corresponds to replacing the Boltzmann factors appearing in
equations (1) and (2) by other Boltzmann factors corresponding to the new
value $T'$, transforming the whole function $P_T(E)$ into $P_{T'}(E)$. In
particular, the probability values are reduced near the former peak, and
enhanced near the new peak position $<E>_{T'}$. However, since the statistics
is poor near this new peak position, the inferred $P_{T'}(E)$ is not
accurate. That is why the histogram method, in spite of its elegant
reasoning, has difficulties in practice [5].

In the present work, we introduce a new method conceived to avoid the
exponential tails responsible for the drawback of the histogram method. It is
presented in the next section. Section III is devoted to show that our
technique is exact for the simple case of the Ising ferromagnet in one
dimension, and to compare our numerical results with the exactly known values
for the same model in two dimensions. Technical details concerning the
computer implementation of the method for the Ising ferromagnet are presented
in section IV (which the reader can skip). Concluding remarks are made in
section V.

\vskip 0.5cm
\centerline{\bf II) The Method}\parG
In this section, we will not restrict ourselves to any particular model.
Suppose one only knows how to compute the energy $E$ of some given state of
the system in the thermodynamic limit (in practice, the system may be a large
but finite one). Our reasoning is based on two steps. First, we define a
non-biased random walk along the $E$ axis. For that purpose, one needs first
to define a set of dynamic movements. For instance, in the case of $N$ Ising
spins one can adopt one-spin flips giving raise to $N$ possible movements
starting from each state. Other dynamics (cluster flips, continuous spin
dynamics, etc.) can also be adopted, depending on the problem being treated.
Within this predefined dynamics, starting from the current state with energy
$E$, consider all possible movements changing this state. These movements can
be classified into two classes~[6]:

\hskip 3cm {class 1:$\,\, E \,\, $\hbox to 30pt {\rightarrowfill}
$\,\, E - \Delta{E}$}\par
\hskip 3cm {class 2:$\,\, E \,\, $\hbox to 30pt {\rightarrowfill}
$\,\, E + \Delta{E}\ \  ,$}\par

\noindent where $\Delta{E} > 0$. Suppose for the moment that all possible
movements have the same $\vert \Delta{E} \vert$. This hypothesis is not
important, and will be disregarded at the end. In the physical region of
positive temperatures, $g(E)$ is a monotonically, fast increasing function of
the energy, and thus the number of possible movements of class 2 is larger
than the corresponding possibilities for class 1. In this way, if one naively
defines the dynamics by performing any randomly tossed movement, the energy
will increase monotonically up to the maximum entropy region corresponding to
infinite temperature. This naive dynamic rule corresponds to a biased random
walk along the energy axis. In order to construct a {\bf non biased} random
walk, we propose the following dynamics: if the currently chosen movement
belongs to class 1, it is accepted and performed; if, however, it belongs to
class 2, it is accepted and performed only with probability $N_{\rm dn} /
N_{\rm up}$, where $N_{\rm dn}$ and $N_{\rm up}$ are the total numbers of
possible movements of classes 1 and 2, respectively, counted at the current
state. This acceptance probability removes the bias, forcing the
probabilities of increasing or decreasing the energy to be equal.  Note that
$N_{\rm dn}$ and $N_{\rm up}$ correspond to the {\bf potential} possible
movements from which only one would be actually performed at each step.
Within this dynamic rule the region already visited along the energy axis
increases its width proportionally to $\Delta{E} \sqrt{t}$, like a random
walk, where $t$ is the number of performed movements, i.e. the length of the
Markovian sequence of states. This makes our method completely distinct from
any other based on the Boltzmann distributions, for which the visited
energies are confined to narrow windows due to the quoted exponential tails.
Figure 1 shows the number of visits as a function of the energy for the Ising
ferromagnet in two dimensions obtained by both methods.

The second step of our reasoning concerns the direct measurement of $g(E)$.
Following the non biased random walk dynamics defined above, the probability
for the energy to jump from $E$ to $E + \Delta{E}$ is the same as that of
jumping back from $E + \Delta{E}$ to $E$. This can be mathematically stated
as

$$<N_{\rm up}(E)> g(E) =\,\, <N_{\rm dn}(E+\Delta{E})>
g(E+\Delta{E})\,\,\,\,\, ,
\eqno(4)$$

\noindent where the averages are again microcanonical. Equation (4) can be
rewritten as

$$\ln g(E+\Delta{E}) - \ln g(E) = \ln {<N_{\rm up}(E)>
\over <N_{\rm dn}(E+\Delta{E})>}
\eqno(5)$$

\noindent or

$$\beta(E) \equiv {{\rm d} \ln g(E) \over {\rm d}E} = {1 \over \Delta{E}}\,\,
\ln {<N_{\rm up}(E)> \over <N_{\rm dn}(E+\Delta{E})>}\,\,\,\,\, ,
\eqno(5)$$

\noindent allowing one to measure $g(E)$ and thus the entropy change, from
the averages $<N_{\rm up}(E)>$ and $<N_{\rm dn}(E)>$ accumulated during the
random walk.

Our method consists, then, in performing the random walk dynamics defined
above, and accumulating values in four histograms along the $E$ axis: the
number of visits; the quantity $Q$ one in interested in; the average number
$N_{\rm up}$ of movements of class 1; and $N_{\rm dn}$ corresponding to class
2. At the end, $g(E)$ is determined by equation (5), and the thermal average
$<Q>_T$ by equations (1), (2) and (3).

As a last remark, let us stress that no thermodynamic concepts are present in
our method. Equations like (1), (2) and (3) involve {\bf two completely
different ingredients}: 1) how the system exchanges energy with the
environment, in the particular present case through the Boltzmann equilibrium
distribution represented by the exponential term $\exp(-E/T)$; and 2) the
system itself, i.e., how its internal energy levels are distributed along the
energy axis, represented here by its spectrum $g(E)$.  Our method concerns
only the latter, i.e., the {\bf signature} of the system, independent of its
environment or thermal exchanges. The thermal average sums whose results are
exemplified in figures 3 and 4 are performed only {\bf after} the Markovian
process was finished, and the spectrum $g(E)$ of the system had already been
determined by the method. In this sense, conceptually, our method is
completely distinct from all other based on thermodynamic grounds. It is just
this independence from thermodynamic constraints which frees us from the
narrow window distributions characteristic of statistical physics, allowing
to obtain rather cheaply the overview through the whole space of states.
Ironically, is just in the study of statistical physics where our method
could become a powerful tool. Another important feature of this
thermodynamic-constraint freedom characteristic of our method is the complete
absence of critical slowing down, distinguishing it from others once more.

\vskip 0.5cm
\centerline{\bf III) Ising Ferromagnet Test}\parG
Consider a ring with $N$ Ising spins pointing up or down. Each pair of
neighboring spins may either be parallel or anti-parallel, the contribution
of this pair to the total energy being zero or one, respectively. In other
words, the total energy $E = 0, 2, 4 \dots$ is the number of broken bonds
(neighboring spins pointing in opposite directions). In this case, the exact
degeneracy can be easily obtained as

$$g_{1D}(E) = 2\,\, {N! \over E! (N-E)!}\,\,\,\,\, ,
\eqno(6)$$

\noindent or

$$\beta(E) = {1 \over 2}\,\, \ln {(N-E)^2 \over E^2}\,\,\,\,\, ,
\eqno(7)$$

\noindent where we have taken the thermodynamic limit $N/2 > E >> 1$.

Let's consider one-spin flips and implement our method in this case.  A
movement will belong to class 1 if the spin to be flipped is currently
surrounded by two broken bonds. On the other hand, class 2 corresponds to
spins parallel to both neighbors. In both cases, the energy jump is
$\Delta{E} = \pm 2$. Neglecting $\Delta{E}$ compared to $E$, one obtains the
averages

$$<N_{\rm dn}(E)>\,\, = {E^2 \over N}
\eqno(8)$$

\noindent corresponding to the probability $(E/N)^2$ of finding two
neighboring broken bonds, and analogously

$$<N_{\rm up}(E)>\,\, = {(N-E)^2 \over N}\,\,\,\,\, .
\eqno(9)$$

\noindent Comparing eqs.~(8) and (9) with eqs. (5) and (7) we see that our
method gives the exact result (6) for the 1D Ising ferromagnet.

The exact degeneracy $g_{2D}(E)$ of the Ising ferromagnet in two dimensions
was also recently derived for finite systems [7], by using the algebraic
software MATHEMATICA, from closed forms already known [8] for finite square
lattices. We use it for another non trivial test of our method. Figure 2
shows the plot of $\ln g(E)$ obtained by our simulation, for a $32 \times 32$
square lattice and the exact curve [7]. They are indistinguishable on the
scale of the plot.

Now, we no longer have the same absolute value $\Delta{E}$ for all possible
one-spin flips. Spins surrounded by zero or four parallel neighbors
correspond to $\Delta{E} = 4$ and belong to classes 1 or 2, respectively.
Analogously, spins surrounded by one or three parallel neighbors correspond
to $\Delta{E} = 2$ and also belong to classes 1 or 2, respectively. One can
adopt two distinct strategies to deal with this feature. First, one can
divide classes 1 and 2 into sub-classes 4 and 2, storing four histograms:
$N^{(4)}_{\rm dn}(E)$, $N^{(2)}_{\rm dn}(E)$, $N^{(2)}_{\rm up}(E)$ and
$N^{(4)}_{\rm up}(E)$ counting spins surrounded by zero, one, three or four
parallel neighbors, respectively. At the end one can measure the degeneracy
$g_{2D}(E)$ by two independent approaches, either using equation (5) with
$<N^{(4)}_{\rm up}(E)>$, $<N^{(4)}_{\rm dn}(E)>$ and $\Delta{E} = 4$, or,
alternatively, with $<N^{(2)}_{\rm up}(E)>$, $<N^{(2)}_{\rm dn}(E)>$ and
$\Delta{E} = 2$. The second strategy corresponds to store only two histograms
for $N_{\rm up}^{1/\Delta{E}}(E)$ and $N_{\rm dn}^{1/\Delta{E}}(E)$,
replacing equation (5) by the equivalent form

$$\beta(E) \equiv {{\rm d} \ln g(E) \over {\rm d}E} =
\ln {<N_{\rm up}(E)^{1/\Delta{E}}> \over
<N_{\rm dn}(E+\Delta{E})^{1/\Delta{E}}>}\,\,\,\,\, .
\eqno(10)$$

\noindent Adopting the first strategy, we confirmed that within the
statistical accuracy, both determinations of $g_{2D}(E)$ give the same
result. They also agree with the values obtained through the second strategy
which is particularly adapted to models where various possible values of
$\Delta{E}$ occur.

Figure 3 shows the averaged energy and specific heat, obtained by the present
method, also indistinguishable from the exact curves [9]. The inset shows the
exact specific heat blowed up near the peak as a continuous line, within the
error bars of our results represented by the crosses. For a larger lattice,
figure 4 shows also the magnetization and susceptibility which can be
compared with canonical Monte Carlo simulations [10], since they are not yet
known exactly for finite lattices. In order to break the global spin flip
symmetry, the magnetization here is considered as the average of the absolute
difference between the population fractions of spins up and down. In both
figures 3 and 4 we considered the energy as twice the number of broken bonds
in order to fit the usual form $-J \sum S_\imath S_\jmath$ of the Ising
Hamiltonian.

\vskip 0.5cm
\centerline{\bf IV) Technical Details}\parG
We have written a C program using some multispin coding tricks [11] in order
to accelerate the code. In particular, we have adopted the multilattice
approach [12] storing the states of 32 lattices in a $L \times L$ square
array of 32-bit integers, where $L$ is the linear size of the lattice. All 32
samples are processed in parallel by using as often as possible bitwise
logical operations instead of algebraic ones [11]. The sites to be flipped
are tossed at random. We adopted the pseudo random number generator which
consists in the multiplication of the current random 32-bit odd unsigned
integer $R$ with 65539, i.e.

$$R = 65539 \times R
\eqno(11)$$

\noindent where only the first (less significant) 32 bits of the result are
keeped. This truncation is automatically done by most compilers. Different
random numbers are tossed for the 32 lattices. We have also adopted periodic
boundary conditions. The starting state is random and then thermalized by 10
entire lattice sweeps at the critical temperature (Metropolis dynamics). The
physical positive-temperature range of energies corresponds to $0 \le e \le
0.5$, where $e$ is the energy per bond. Since the Onsager critical energy
corresponds to $e \approx 0.146$, we decided to restrict our random walk
energy range to $0 < e < 0.4$, tossing a new initial state every time the
current energy goes out of this range. All our numerical results correspond
to the kind of calculations described in this paragraph (except for data
shown in figures 1 and 2, for which we have extended the range up to $e =
0.5$).

Implementing our method as described above, the correlations between
successive states decay slowly: around 10 lattice sweeps must be taken
between sucessive data picked for latter averages, in order to obtain good,
correlation-free results. Another characteristic feature of Ising one-spin
flip dynamics is the frequent blocking of the system in certain pathological
low energy states [13]. If for instance, the square lattice presents a
vertical strip of up spins in adjacent columns, all other spins pointing
down, the Markovian sequence will be blocked in this state. This does not
change the results, because we only store values to perform the averages if
the current state differs from the previous one, but computer time is wasted
by these blockings.

In order to accelerate the code, we decided to overcome both problems quoted
in the previous paragraph. To do this, we introduced 5 (non-averaging)
lattice sweeps of the Metropolis dynamics after each lattice sweep using our
random walk dynamics. The fixed temperature adopted for these 5 extra
thermalization sweeps corresponds to that of the current energy as measured
by equation (10) (note that eq.~(10) is formally equal to the statistical
definition of the inverse temperature), extracting the averages from the
values already accumulated in the histograms $N_{\rm up}^{1/\Delta{E}}$ and
$N_{\rm dn}^{1/\Delta{E}}$. In order to improve efficiency even more one can
use the ratio between the corresponding values already accumulated in the
histograms instead of the current instantaneous ratio between the numbers of
class 1 and class 2 spins.

As a result of all these acceleration tricks the total computer time is less
than 40 minutes on a PENTIUM PC running at 66MHz frequency, for $(1+5) \times
10^4$ lattice sweeps for 32 samples of size $32 \times 32$. The $64 \times
64$ lattice simulation is 4 times slower, and so on. These times are only
$50\%$ larger than we measured for the histogram method under the same
conditions.

\vskip 0.5cm
\centerline{\bf Conclusions}\parG

We have presented a new histogram Monte Carlo method which as compared to the
traditional one based on temperature [3,4] is based on histograms measured
from a random walk along the energy axis. These histograms have the advantage
of having much broader tails allowing to extrapolate to a much larger range
of temperatures with a rather small number of samples. We have tested our
method on the two-dimensional Ising model and succeeded in reproducing
thermodynamic quantities with high accuracy over the entire physical
temperature scale with very little effort.

Our method is very general and could be useful for instance for simulations
of spin glasses or spin models in three dimensions. Work in this direction
is in progress.

\vskip 0.5cm
\centerline{\bf Acknowlegments}\parG

We thank D. Stauffer, S. Moss de Oliveira, C. Moukarzel, Yi-Cheng Zhang and
D.P. Landau for helpful discussions. One of us (PMCO) thanks also for the
hospitality of Geraldo and \^Angela S\'a at B\'uzios, Rio de Janeiro, where
most of this work was done.

\vskip 0.5cm
\centerline{\bf References}\parG
\item{[1]} Binder K. (ed.), {\sl Monte Carlo Methods in Statistical Physics},
Topics in Current Physics vols. 7, 36 and 71 (Springer, 1986).\par
\item{[2]} Metropolis N., Rosenbluth A.W., Rosenbluth M.N., Teller A.H. and
Teller E., {\it J. Chem. Phys.} {\bf 21}, 1087 (1953).\par
\item{[3]} Salzburg Z.W., Jacobson J.D., Fickett W. and Wood W.W., {\it J.
Chem. Phys.} {\bf 30} 65 (1959).\par
\item{[4]} Ferrenberg A.M. and Swendsen R.H., {\it Phys. Rev. Lett.} {\bf 61},
2635 (1988).\par
\item{[5]} Ferrenberg A.M. and Landau D.P., {\it Phys. Rev.} {\bf B44}, 5081
(1991).\par
\item{[6]} One can also have a third class of movements conserving the energy
which does not influence our reasoning. Nevertheless, these extra movements
can be performed in order to accelerate the decorrelation between successive
members of the Markovian sequence of states. We have adopted this strategy in
all our numerical simulations.\par
\item{[7]} Beale P.D., {\it Phys. Rev. Lett.} {\bf 76}, 78 (1996).\par
\item{[8]} Kauffmann B., {\it Phys. Rev.} {\bf 76}, 1232 (1949), after the
solution for infinity lattices by Onsager L., {\it Phys. Rev.} {\bf 65}, 117
(1944).\par
\item{[9]} Ferdinand A.E. and Fisher M.E., {\it Phys. Rev.} {\bf 185}, 832
(1969).\par
\item{[10]} see, for instance, Landau D.P., {\it Phys. Rev.} {\bf B13}, 2997
(1976), or de Oliveira P.M.C. and Penna T.J.P., {\it Rev.  Bras. F\'\i s.}
{\bf 18}, 502 (1988).\par
\item{[11]} de Oliveira P.M.C., {\sl Computing Boolean Statistical Models}
World Scientific, Singapore, ISBN 981-02-0238-5 (1991).\par
\item{[12]} Bhanot G., Duke D. and Salvador R., {\it J. Stat. Phys.} {\bf
44}, 985 (1986); {\it Phys. Rev.} {\bf B33}, 7841 (1986).\par
\item{[13]} see, for instance, Derrida B., de Oliveira P.M.C. and Stauffer
D., {\it Physica} {\bf A224}, 604 (1996).\par

\vfill\eject
\centerline{\bf Figure Captions}\parG
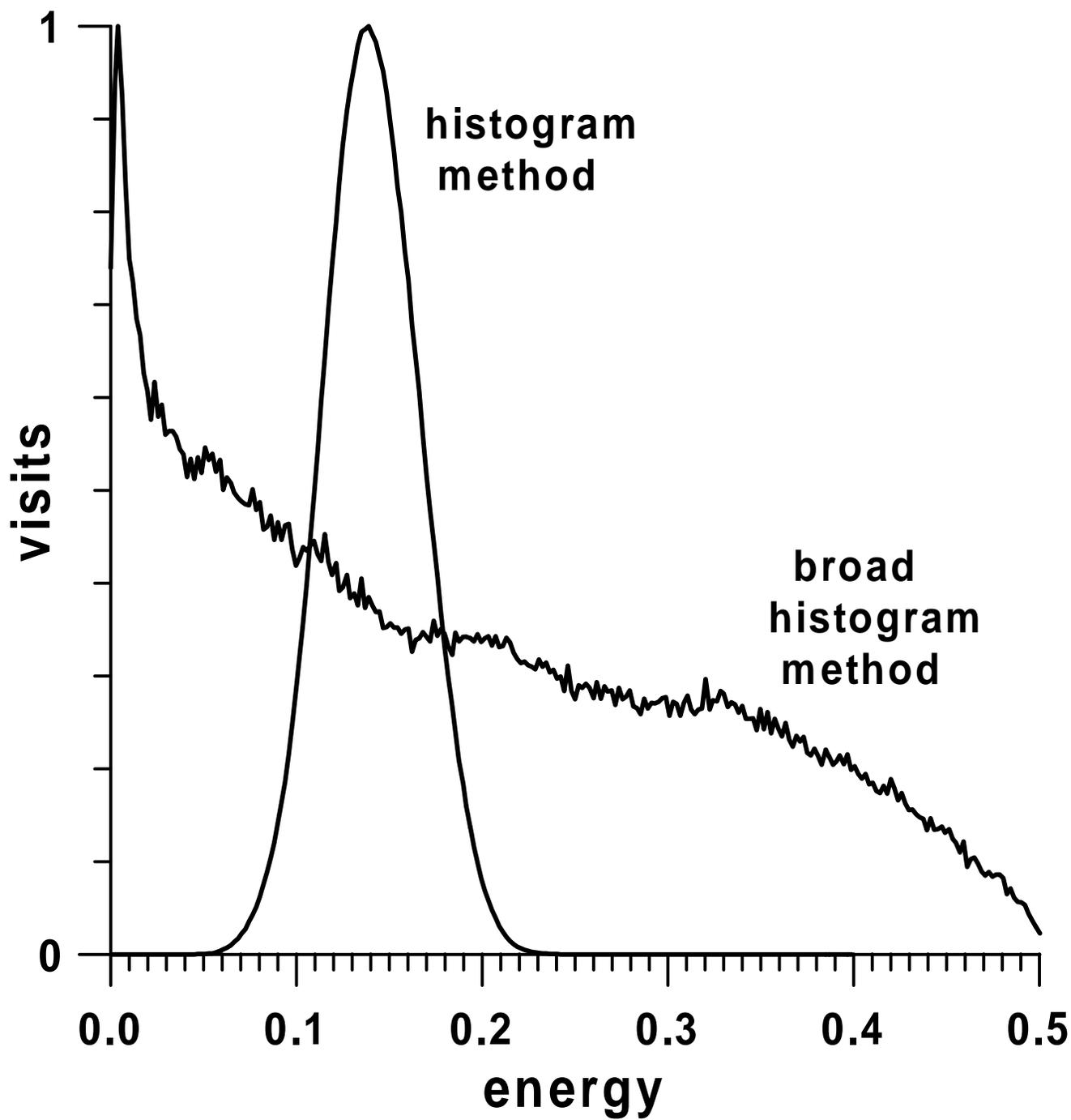
\item{figure 1} Number of visits as a function of energy obtained from the
histogram method [3,4] (fixing the temperature at the critical value), and
from the present method, for the Ising ferromagnet on a $32 \times 32$ square
lattice. The energy values are displayed by the density of broken bonds. On
that scale the whole physical positive-temperature range is between $0$
(ground state) and $0.5$ (infinite temperature), while the critical point
corresponds to $0.146$. Within the present method, the whole physical energy
range is explored for any system size. In contrast, the histogram method
explores only a narrow window, the larger the system size the narrower is
this window: namely, for $L = 32$ (shown in the figure), $64$, $128$ and
$256$, we found widths $0.061$, $0.031$, $0.015$ and $0.007$,
respectively.\par

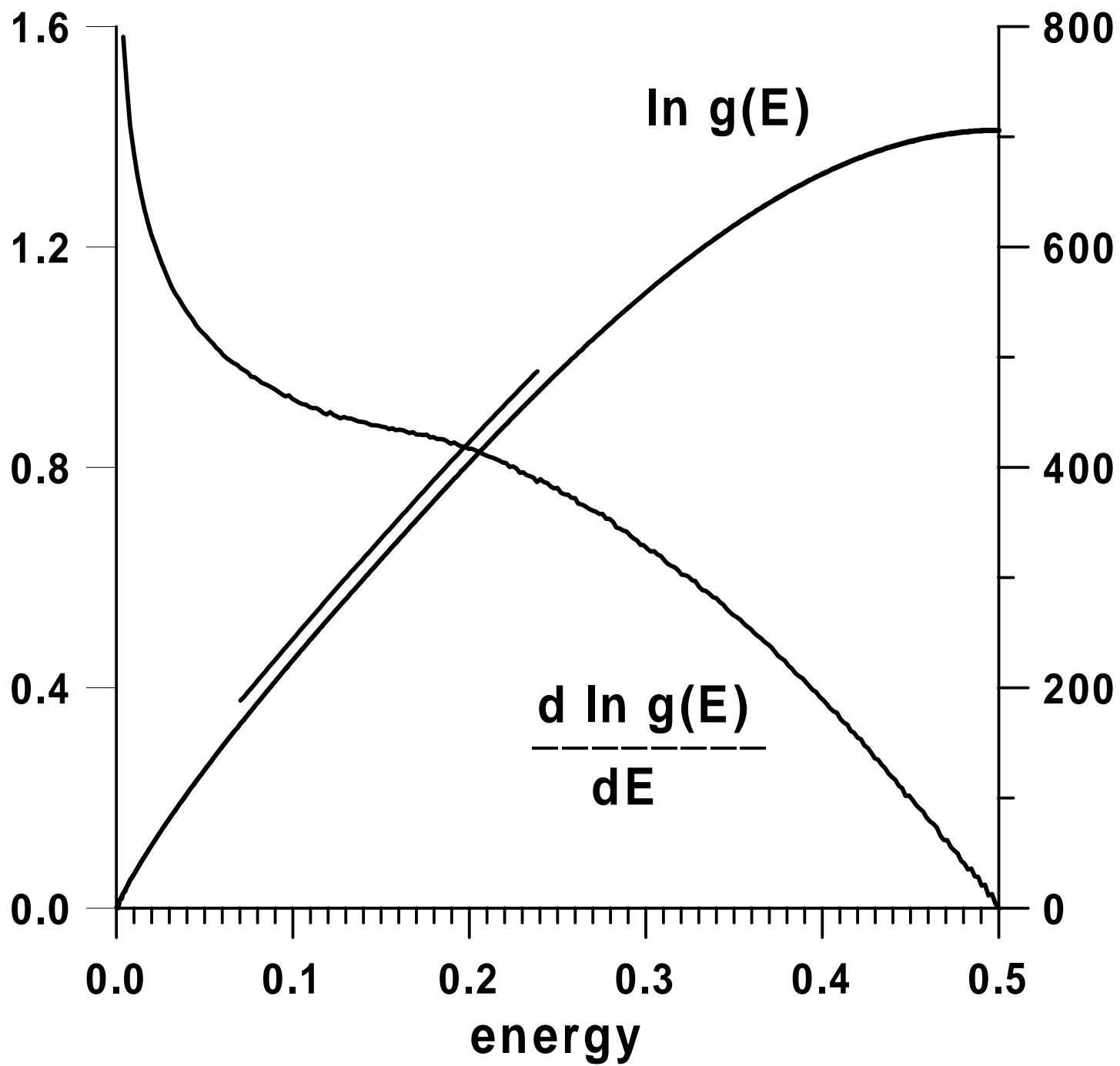
\item{figure 2} Degeneracy $g(E)$ for the Ising ferromagnet on a $32 \times
32$ square lattice (the exact result [8] could be explicitly obtained up to
this size, through an algebraic MATHEMATICA program [7]). The energy values
are displayed by the density of broken bonds. The plot contains both the
exact curve and the results of our simulations and they are indistinguishable
at this scale. The shorter line displayed slightly above corresponds to the
histogram method [3,4] for which the results remain restricted inside the
narrow windows mentioned on the caption of figure 1.  Also shown is the
derivative which is the quantity we directly measured.\par

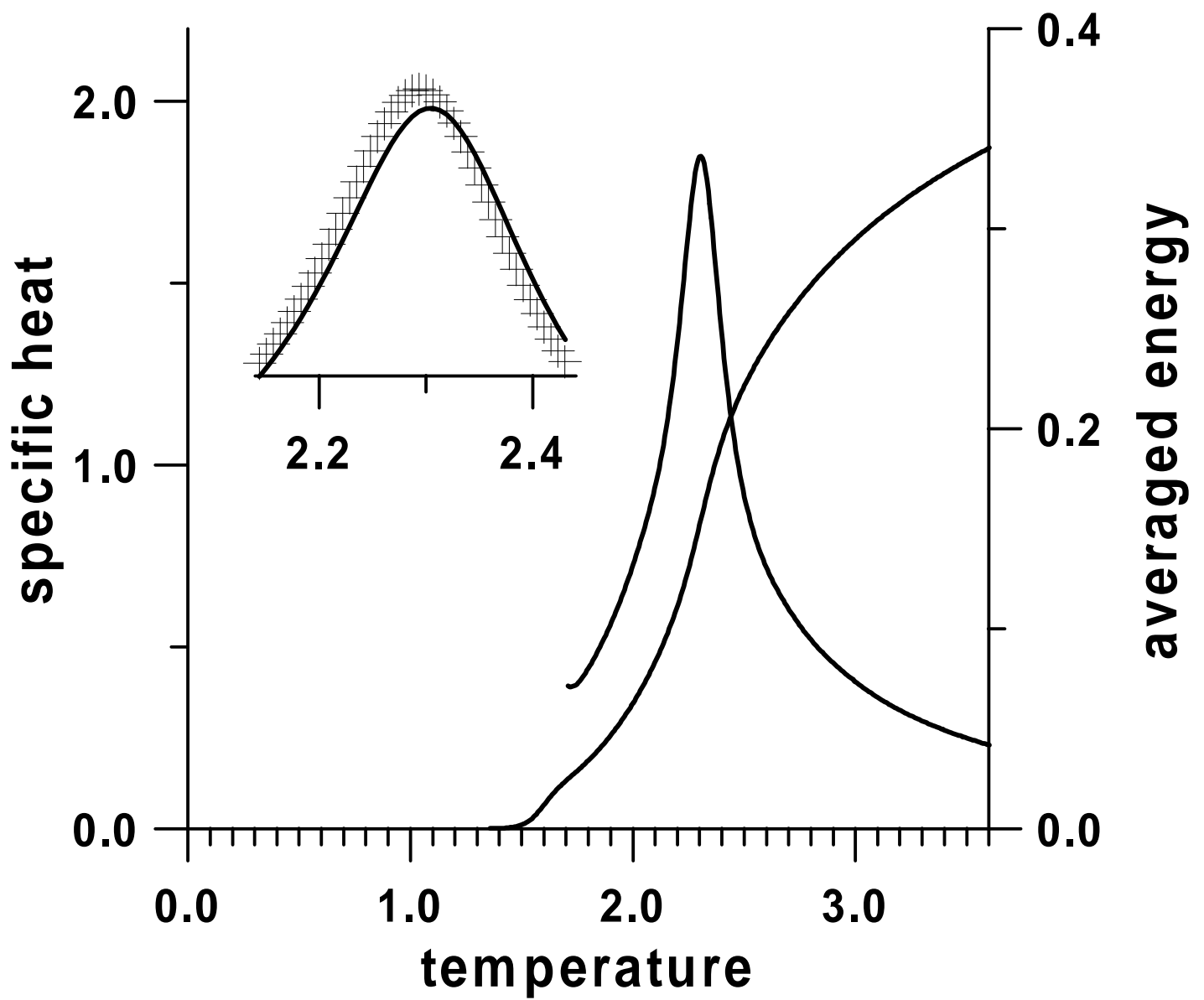
\item{figure 3} Averaged energy and specific heat obtained from the present
method for the Ising ferromagnet on a $32 \times 32$ square lattice. The
inset compares the exactly known curve [9] with our results (symbols of the
same size of the error bars), near the specific heat peak: these curves are
indistinguishable at the larger scale. Concerning CPU time, we took less than
40 minutes on a PENTIUM PC running at 66Mhz, $50\%$ more than the histogram
method at the same conditions.\par

\item{figure 4} Curves obtained from the present method for the Ising
ferromagnet on a $128 \times 128$ square lattice. Only $10^4$ whole lattice
sweeps are used, the same amount adopted for the smaller lattices in previous
figures.\par
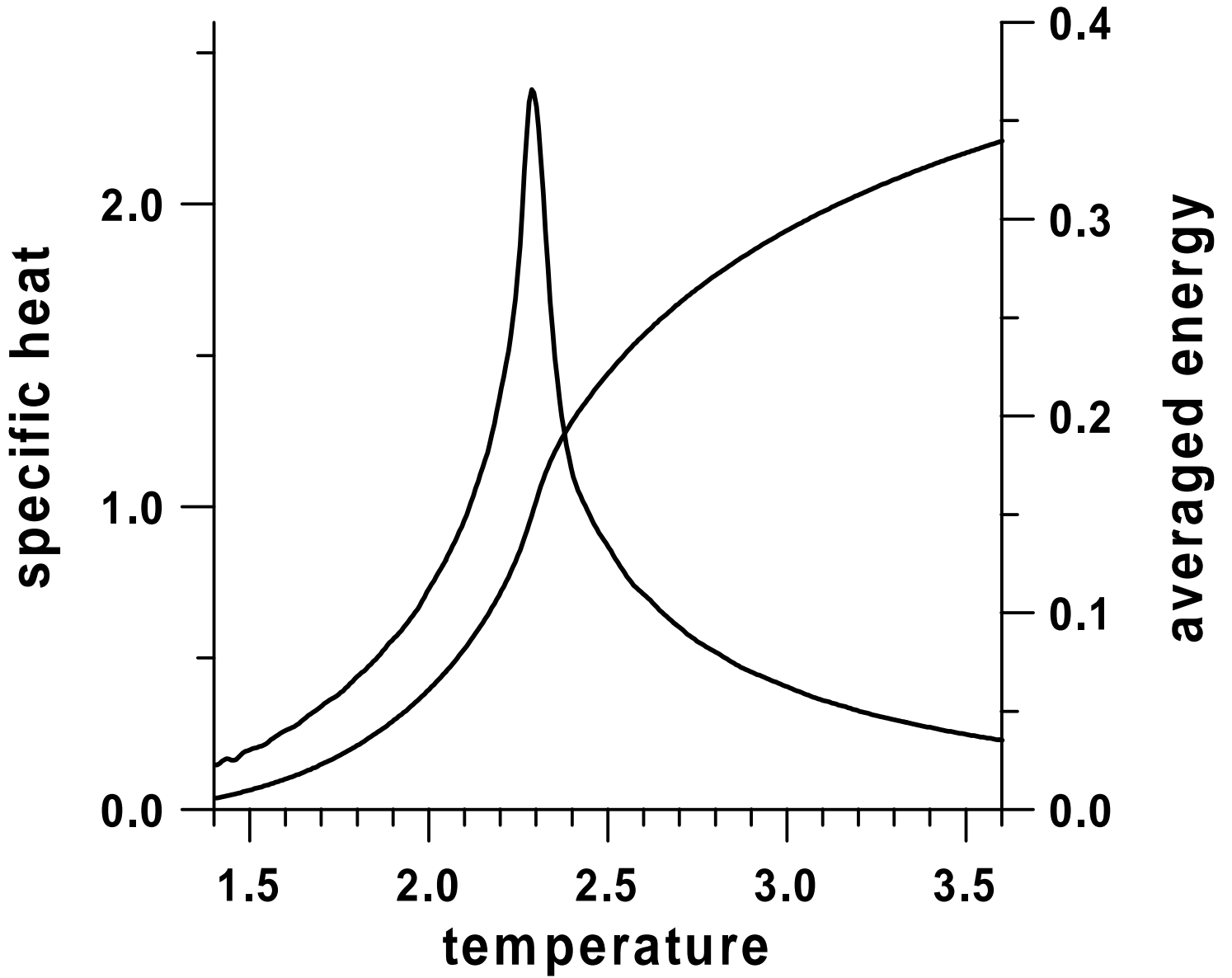
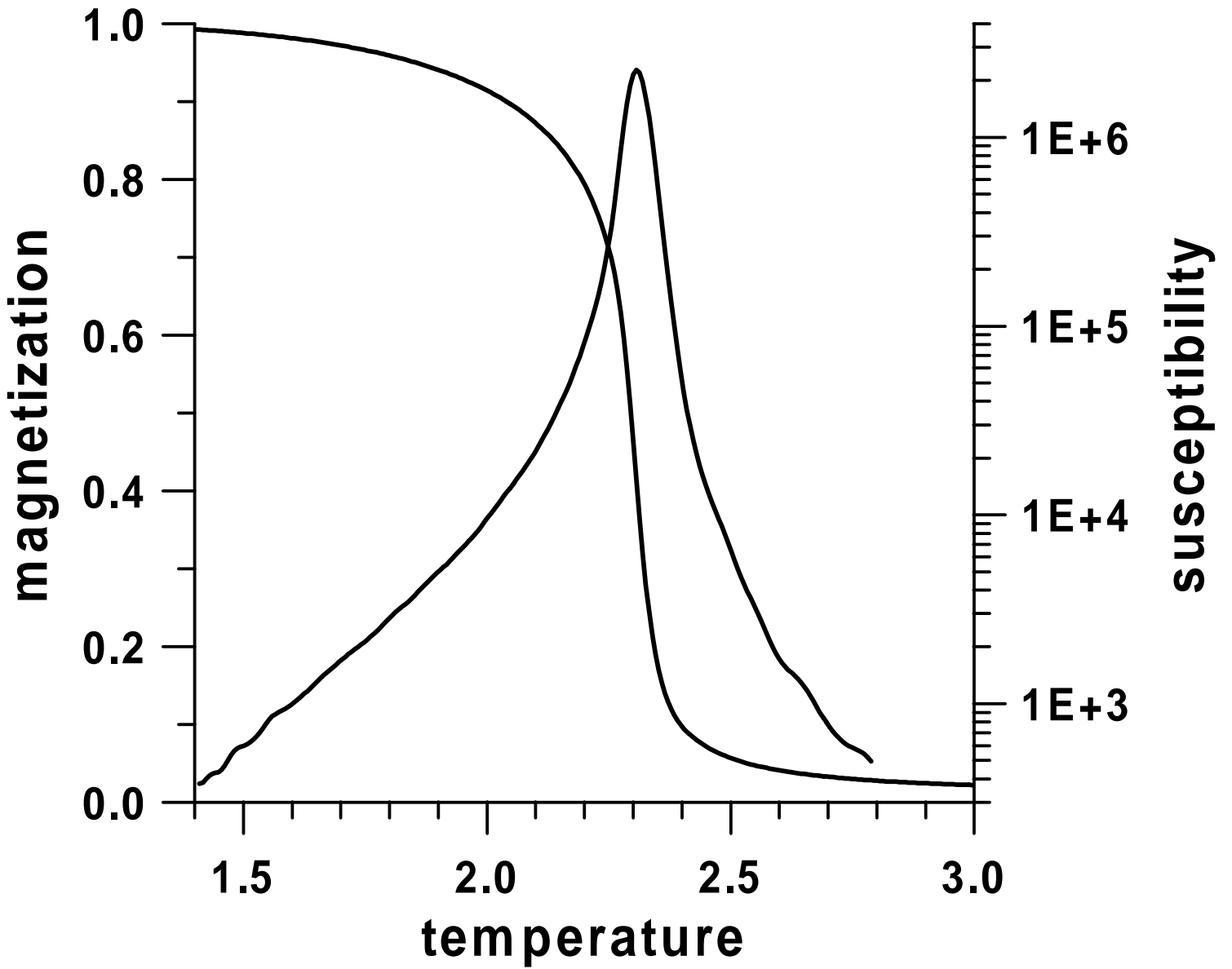
\bye